\documentclass{llncs}
\usepackage[T1]{fontenc}
\usepackage[utf8]{inputenc}
\usepackage[bookmarks=true,hidelinks,colorlinks=true,allcolors=blue]{hyperref}
\usepackage{amssymb}
\usepackage{todonotes}

\usepackage[american]{babel}
\usepackage{graphicx}
\usepackage{caption}
\usepackage{subcaption}

\usepackage{isabelle}
\usepackage{isabellesym}
\isabellestyle{it}
\def\isastyle{\fontfamily{lmss}\selectfont}
\def\isastyleminor{\fontfamily{pcr}\selectfont}

\newcommand{\DefineSnippet}[2]{\expandafter\newcommand\csname snippet--#1\endcsname{\begin{isabellebody}#2\end{isabellebody}}}
\newcommand{\Snippet}[1]{
  \csname snippet--#1\endcsname}
\DefineSnippet{strengthenpre}{
\isacommand{lemma}\isamarkupfalse%
\ strengthen{\isacharunderscore}pre{\isacharcolon}\isanewline
\ \ {\isachardoublequoteopen}{\isasymlbrakk}\ {\isasymforall}s{\isachardot}\ P'\ s\ {\isasymlongrightarrow}\ P\ s{\isacharsemicolon}\ \ {\isasymturnstile}\isactrlsub t\ {\isacharbraceleft}P{\isacharbraceright}\ c\ {\isacharbraceleft}Q{\isacharbraceright}\ {\isasymrbrakk}\ {\isasymLongrightarrow}\ {\isasymturnstile}\isactrlsub t\ {\isacharbraceleft}P'{\isacharbraceright}\ c\ {\isacharbraceleft}Q{\isacharbraceright}{\isachardoublequoteclose}\isanewline
\ \ %
\isacommand{by}\isamarkupfalse%
\ {\isacharparenleft}metis\ conseq{\isacharparenright}%
{\isafoldproof}%
}%EndSnippet
\DefineSnippet{whilefun}{
\isacommand{lemma}\isamarkupfalse%
\ While{\isacharunderscore}fun{\isacharcolon}\isanewline
\ \ {\isachardoublequoteopen}{\isasymlbrakk}\ {\isasymAnd}n{\isacharcolon}{\isacharcolon}nat{\isachardot}\ {\isasymturnstile}\isactrlsub t\ {\isacharbraceleft}{\isasymlambda}s{\isachardot}\ P\ s\ {\isasymand}\ bval\ b\ s\ {\isasymand}\ n\ {\isacharequal}\ f\ s{\isacharbraceright}\ c\ {\isacharbraceleft}{\isasymlambda}s{\isachardot}\ P\ s\ {\isasymand}\ f\ s\ {\isacharless}\ n{\isacharbraceright}{\isasymrbrakk}\isanewline
\ \ \ {\isasymLongrightarrow}\ {\isasymturnstile}\isactrlsub t\ {\isacharbraceleft}P{\isacharbraceright}\ WHILE\ b\ DO\ c\ {\isacharbraceleft}{\isasymlambda}s{\isachardot}\ P\ s\ {\isasymand}\ {\isasymnot}bval\ b\ s{\isacharbraceright}{\isachardoublequoteclose}\isanewline
\ \ %
\isacommand{by}\isamarkupfalse%
\ {\isacharparenleft}rule\ Hoare{\isacharunderscore}Total{\isachardot}While\ {\isacharbrackleft}\isakeyword{where}\ T{\isacharequal}{\isachardoublequoteopen}{\isasymlambda}s\ n{\isachardot}\ n\ {\isacharequal}\ f\ s{\isachardoublequoteclose},\ simplified{\isacharbrackright}{\isacharparenright}%
{\isafoldproof}%
}%EndSnippet
\DefineSnippet{execexec}{
\isacommand{lemma}\isamarkupfalse%
\ exec{\isacharunderscore}n{\isacharunderscore}exec{\isacharcolon}\isanewline
\ \ {\isachardoublequoteopen}P\ {\isasymturnstile}\ c\ {\isasymrightarrow}{\isacharcircum}n\ c\ {\isasymLongrightarrow}\ P\ {\isasymturnstile}\ c\ {\isasymrightarrow}{\isacharasterisk}\ c{\isachardoublequoteclose}\isanewline
\ \ %
\isacommand{by}\isamarkupfalse%
\ {\isacharparenleft}induct\ n\ arbitrary{\isacharcolon}\ c{\isacharparenright}\ {\isacharparenleft}auto\ intro{\isacharcolon}\ star{\isachardot}step{\isacharparenright}%
{\isafoldproof}%
\isanewline
\isanewline
\isanewline
\isanewline
\isanewline
\isanewline
\isacommand{lemma}\isamarkupfalse%
\ exec{\isacharunderscore}exec{\isacharunderscore}n{\isacharcolon}\isanewline
\ \ {\isachardoublequoteopen}P\ {\isasymturnstile}\ c\ {\isasymrightarrow}{\isacharasterisk}\ c\ {\isasymLongrightarrow}\ {\isasymexists}n{\isachardot}\ P\ {\isasymturnstile}\ c\ {\isasymrightarrow}{\isacharcircum}n\ c{\isachardoublequoteclose}\isanewline
\ \ %
\isacommand{by}\isamarkupfalse%
\ {\isacharparenleft}induct\ rule{\isacharcolon}\ star{\isachardot}induct{\isacharparenright}\ {\isacharparenleft}auto\ intro{\isacharcolon}\ exec{\isacharunderscore}Suc{\isacharparenright}%
{\isafoldproof}%
\isanewline
\isanewline
\isanewline
\isacommand{lemma}\isamarkupfalse%
\ exec{\isacharunderscore}eq{\isacharunderscore}exec{\isacharunderscore}n{\isacharcolon}\isanewline
\ \ {\isachardoublequoteopen}{\isacharparenleft}P\ {\isasymturnstile}\ c\ {\isasymrightarrow}{\isacharasterisk}\ c{\isacharparenright}\ {\isacharequal}\ {\isacharparenleft}{\isasymexists}n{\isachardot}\ P\ {\isasymturnstile}\ c\ {\isasymrightarrow}{\isacharcircum}n\ c{\isacharparenright}{\isachardoublequoteclose}\isanewline
\ \ %
\isacommand{by}\isamarkupfalse%
\ {\isacharparenleft}blast\ intro{\isacharcolon}\ exec{\isacharunderscore}exec{\isacharunderscore}n\ exec{\isacharunderscore}n{\isacharunderscore}exec{\isacharparenright}%
{\isafoldproof}%
}%EndSnippet
\DefineSnippet{smalltobig}{
\isacommand{lemma}\isamarkupfalse%
\ small{\isacharunderscore}to{\isacharunderscore}big{\isacharcolon}\isanewline
\ \ {\isachardoublequoteopen}cs\ {\isasymrightarrow}{\isacharasterisk}\ {\isacharparenleft}SKIP,t{\isacharparenright}\ {\isasymLongrightarrow}\ cs\ {\isasymRightarrow}\ t{\isachardoublequoteclose}\isanewline
\isacommand{apply}\isamarkupfalse%
\ {\isacharparenleft}induction\ cs\ {\isachardoublequoteopen}{\isacharparenleft}SKIP,t{\isacharparenright}{\isachardoublequoteclose}\ rule{\isacharcolon}\ star{\isachardot}induct{\isacharparenright}\isanewline
\isacommand{apply}\isamarkupfalse%
\ {\isacharparenleft}auto\ intro{\isacharcolon}\ small1{\isacharunderscore}big{\isacharunderscore}continue{\isacharparenright}\isanewline
\isacommand{done}\isamarkupfalse%
{\isafoldproof}%
}%EndSnippet

\usepackage{framed}
\usepackage{bbm}
\definecolor{ltblue}{rgb}{0,0.4,0.4}
\definecolor{dkblue}{rgb}{0,0.1,0.6}
\definecolor{dkgreen}{rgb}{0,0.35,0}
\definecolor{dkgreen}{rgb}{0,0.35,0}
\definecolor{dkviolet}{rgb}{0.3,0,0.5}
\definecolor{dkred}{rgb}{0.5,0,0}
\definecolor{orange}{rgb}{0.9,0.5,0.3}
\definecolor{violet}{rgb}{0.7,0,0.7}
\usepackage{listings}

\renewcommand{\tt}{\theta}

\newcommand\pca[3]{\{#1\}\kern1pt{#2}\kern1pt\{#3\}}

\usepackage{mathrsfs}

\definecolor{colorsec}{HTML}{345A8A}
\definecolor{colorsubsec}{HTML}{4F81BD}
\definecolor{colorsubsubsec}{HTML}{5388C8}
\hypersetup{
  colorlinks = true,
  citecolor=purple,
  linkcolor=colorsubsec,
}

\definecolor{bg}{rgb}{0.95,0.95,0.95}

\makeatletter

\usepackage[framemethod=tikz]{mdframed}
\usepackage[most]{tcolorbox}

\def\beginlstdelim#1#2#3#4%
{%
  \def\endlstdelim{#2\egroup}%
  {\ttfamily#3#1}\bgroup#4\aftergroup\endlstdelim%
}

\lstdefinelanguage{smt}{
  language=lisp,
  alsoletter=0123456789>=,
  keywords={define-fun,declare-fun,declare-const,set-option,echo,set-info,exit,pop,push,assert},
  classoffset=1,
  morekeywords={Int,Real,Bool},keywordstyle=\color{blue!60!black}\bfseries,
  classoffset=2,
  morekeywords={not,and,or,=>},keywordstyle=\bfseries,
  classoffset=3,
  morekeywords={check-sat,check-sat-assume},keywordstyle=\color{red}\bfseries,
  classoffset=4,
  morekeywords={true,false,0,1,2,3,4,5,6,7,8,9,10},keywordstyle=\color{violet}\bfseries,
  classoffset=0,
  moredelim=**[is][\beginlstdelim{define-fun\ }{\ }{\color{green!50!black}\bfseries}{\color{blue!80!black!50!white}\bfseries}]{define-fun\ }{\ },
  moredelim=**[is][\beginlstdelim{declare-const\ }{\ }{\color{green!50!black}\bfseries}{\color{blue!80!black!50!white}\bfseries}]{declare-const\ }{\ },
  moredelim=**[is][\beginlstdelim{declare-fun\ }{\ }{\color{green!50!black}\bfseries}{\color{blue!80!black!50!white}\bfseries}]{declare-fun\ }{\ },
  mathescape=false
}

\lstdefinelanguage{Coq}{
  alsoletter={<}{:}0123456789,
  morekeywords={Variable,Inductive,CoInductive,Fixpoint,CoFixpoint,%
    Definition,Program, Lemma,Theorem,Corollary,Axiom,Local,Save,Grammar,Syntax,Intro,%
    Trivial,Qed,Intros,Symmetry,Simpl,Rewrite,Apply,Elim,Assumption,%
    Left,Cut,Case,Auto,Unfold,Exact,Right,Hypothesis,Pattern,Destruct,%
    Constructor,Defined,Fix,Record,Proof,Induction,Hints,Exists,%
    Parameter,Parameters,Split,Red,Reflexivity,Transitivity,if,then,else,Opaque,Module,%
    Transparent,Inversion,Absurd,Generalize,Mutual,Cases,of,Analyze,%
    AutoRewrite,Functional,Scheme,params,Refine,using,Discriminate,Try,%
    Require,Load,Import,Scope,Open,Section,End,Ltac,fun,forall,exists,Canonical,Structure,Eval,Notation,as,return,Goal,Class,Module%
  },%
  classoffset=1,
  morekeywords={Type,Prop,bool,nat,Set,let,in,match,with,end,as,<:,Z,farray,bitvector},keywordstyle=\color{blue!60!black}\bfseries,
  classoffset=2,
  morekeywords={Error:,Warning:},keywordstyle=\color{red}\bfseries,
  classoffset=3,
  morekeywords={0,1,2,3,4,5,6,7,8,9,10,11,12,13,14,15,16,16,18,19,20},keywordstyle=\color{violet},
  classoffset=0,
  sensitive, %
  moredelim=**[is][\beginlstdelim{Inductive\ }{\ }{\color{green!50!black}\bfseries}{\color{blue!80!black!50!white}\bfseries}]{Inductive\ }{\ },
  moredelim=**[is][\beginlstdelim{Definition\ }{\ }{\color{green!50!black}\bfseries}{\color{blue!80!black!50!white}\bfseries}]{Definition\ }{\ },
  moredelim=**[is][\beginlstdelim{Lemma\ }{\ }{\color{green!50!black}\bfseries}{\color{blue!80!black!50!white}\bfseries}]{Lemma\ }{\ },
  moredelim=**[is][\beginlstdelim{Axiom\ }{\ }{\color{green!50!black}\bfseries}{\color{blue!80!black!50!white}\bfseries}]{Axiom\ }{\ },
  moredelim=**[is][\beginlstdelim{Theorem\ }{\ }{\color{green!50!black}\bfseries}{\color{blue!80!black!50!white}\bfseries}]{Theorem\ }{\ },
  moredelim=**[is][\beginlstdelim{Class\ }{\ }{\color{green!50!black}\bfseries}{\color{blue!60!black}\bfseries}]{Class\ }{\ },
  moredelim=**[is][\beginlstdelim{Module\ }{\ }{\color{green!50!black}\bfseries}{\color{blue!60!black}\bfseries}]{Module\ }{\ },
  moredelim=**[is][\beginlstdelim{Record\ }{\ }{\color{green!50!black}\bfseries}{\color{blue!60!black}\bfseries}]{Record\ }{\ },
  morecomment=[n]{(*}{*)},%
  morestring=[d]",%
  literate=
  {=>}{{$\Rightarrow$}}1
  {==>}{{$\Longrightarrow$}}1
  {->}{{$\,\to\,$}}1
  {<-}{{$\leftarrow$}}1
  {>->}{{$\rightarrowtail$}}2
  {*->}{{$\leftrightarrow$}}1
  {forall}{{\color{blue!60!black}\bfseries$\forall$}}1
  {exists}{{\color{blue!60!black}\bfseries$\exists$}}1
  {//n}{{$\neq$}}1
  {<=}{{$\overset{?}{\leq}$}}1
  {>=}{{$\overset{?}{\geq}$}}1  
  {>}{{$\overset{?}{>}$}}1
   {<}{{$<$}}1
  {:=}{{$\triangleq$}}1
  {\/\\}{{$\wedge$}}1
  {|-}{{$\vdash$}}1
  {\\\/}{{$\vee$}}1
  {'}{'}1
  {⟦}{{$\llbracket$}}1
  {⟧}{{$\rrbracket$}}1
  {-->}{{$\longrightarrow$}}1
}

\lstdefinelanguage{lfsc}{
  language=lisp,
    alsoletter={!}{\%}{@}{\\},
  keywords={check,define,declare,program},
  classoffset=1,
  morekeywords={int,mpz,th_holds,holds,term,sort,type,match,fail,default},keywordstyle=\color{blue!60!black}\bfseries,
  classoffset=2,
  keywords={\%,@,!,\\},keywordstyle=\color{violet}\bfseries,
  classoffset=0,
  moredelim=**[is][\beginlstdelim{define\ }{\ }{\color{green!50!black}\bfseries}{\color{green!50!black!50!white}\bfseries}]{define\ }{\ },
  moredelim=**[is][\beginlstdelim{declare\ }{\ }{\color{green!50!black}\bfseries}{\color{blue!80!black!50!white}\bfseries}]{declare\ }{\ },
  moredelim=**[is][\beginlstdelim{program\ }{\ }{\color{green!50!black}\bfseries}{\color{blue!80!black!50!white}\bfseries}]{program\ }{\ },
  escapechar=\&
}

\lstdefinelanguage{smtcoq}{
  alsoletter=\#0123456789\=,
  classoffset=0,
  keywords={or,and,not,impl,true,false,\=,->},keywordstyle=\color{black}\bfseries,
  classoffset=1,
  morekeywords={0,1,2,3,4,5,6,7,8,9,10,11,12,13,14,15,16,16,18,19,20},keywordstyle=\color{violet}\bfseries,
  classoffset=0,
  sensitive=true,
  moredelim=**[is][\beginlstdelim{:(}{\ }{}{\color{green!50!black}\bfseries}]{:(}{\ },
}

\lstdefinelanguage{ocaml}{
  language=[Objective]caml,
  identifierstyle=\ocidstyle
}

\newcommand*\ocidstyle{%
        \expandafter\id@style\the\lst@token\relax
}
\def\id@style#1#2\relax{%
        \ifcat#1\relax\else
                \ifnum`#1=\uccode`#1%
                        \color{blue!60!black}
                \fi
        \fi
}

\newenvironment{tcb}[2][\small]{%
  \tcblisting{enhanced jigsaw,breakable,lines before break=3,
    listing only,colback=bg,colframe=bg,enlarge
    top by=0mm,top=0pt,bottom=0pt,left=2pt,right=2pt,enhanced,
    before={\vspace{10pt}},
    after={\par\vspace{5pt}\noindent},
    listing options={language=#2,basicstyle={\ttfamily#1\upshape}}%
    }}{\endtcblisting}

\newcommand{\code}[1]{\lstinline!#1!}

\makeatother

\begin{document}

\author{Łukasz Czajka\inst{1} \and Burak Ekici\inst{2} \and Cezary Kaliszyk\inst{2}\orcidID{{0000-0002-8273-6059}}}
\title{Concrete Semantics with Coq and CoqHammer}

\institute{
University of Copenhagen, Copenhagen, Denmark\\
\url{luta@di.ku.dk}
\and University of Innsbruck, Innsbruck, Austria\\
\url{burak.ekici,cezary.kaliszyk@uibk.ac.at}
}
\maketitle{}

This is an author's version of a paper published at CICM
2018. Copyright by Springer. The final authenticated publication is
available online at \url{https://doi.org/10.1007/978-3-319-96812-4_5}.

\begin{abstract}
  The ``Concrete Semantics'' book gives an introduction to imperative
  programming languages accompanied by an Isabelle/HOL formalization.
  In this paper we discuss a re-formalization of the book using the
  Coq proof assistant. In order to achieve a similar brevity of the
  formal text we extensively use CoqHammer\footnote{release:
    \url{https://github.com/lukaszcz/coqhammer/releases/tag/v1.0.8-coq8.7}},
  as well as Coq Ltac-level automation. We compare the formalization
  efficiency, compactness, and the readability of the proof scripts
  originating from a Coq re-formalization of two chapters from the
  book.
\end{abstract}

\section{Introduction}
Formal proofs allow today most precise descriptions and specifications of computer
systems and programs. Such precision is very important both for human learning and for
machine knowledge management. Formalization accompanied courses allow students to
investigate the topic to an arbitrary level of detail, and naturally offer very
precise exercises of the topic~\cite{NipkowKlein}. Formalization attached to
mathematical knowledge allows algorithms to the knowledge semantically and permits
learning machine translation to, from, and between datasets~\cite{KaliszykUVG14}. This
becomes even more important with multi-translation, where the availability of the
same text in multiple languages improves the computer-understanding and ability to
translate between each two~\cite{DongWHYW15}.

In this short paper we translate parts of the Concrete Semantics book by Nipkow and
Klein to Coq. To do so, we improve the CoqHammer~{\cite{lcck-jar18}} automation
to be able to handle the more advanced use-cases, improve the legibility of the
reconstructed proofs and compare the proof style and other differences in between
the two. The project is in some ways similar to the ``Certified Programming
with Dependent Types'' book \cite{Chlipala2013}, however we attempt to avoid
dependent types and more advanced constructions to build both an easier material
for students and a more precise dataset for bootstrapping an automated translation
between proof corpora in the style of~\cite{KaliszykUVG14}.

\section{Concrete Semantics with Isabelle/HOL}
The Concrete Semantics book~\cite{NipkowKlein} by Nipkow and Klein is made of two parts.
The first part introduces how to write functional programs, inductive definitions and
how to reason about their properties in Isabelle/HOL’s structured proof language.
While the second part is devoted to formal semantics of programming languages
using the ``small'' imperative \texttt{IMP}\footnote{\texttt{IMP} is a standard
Turing complete imperative language involving the mutable global state as a computational side
effect.
The reason why this language has
been selected is just that it has enough expressive power to be Turing complete.}
language as the instance.
This part more concretely examines several topics in a wide range varying
from operational semantics, compiler correctness to Hoare Logic. The proofs presented in this
part are not given in Isabelle/HOL's structured language. However, such a formalization
accompanies the paper proofs via the provided links usually given in section beginnings.

In this work we attempt to reformalize in Coq some subset of the Isabelle/HOL theories
that accompanies the second part of the book. As illustrated in Chapter~\ref{castd}, 
we aim at catching the same level of automation in Coq thus
approximating the proof texts to the original ones in terms of length.
To do so, we use automated reasoning techniques discussed in Chapter~\ref{coqa}.

%{Half-a-page outline of Part II of the book}
\section{Coq and Coq Automation}\label{coqa}

The Coq proof assistant is based on the Calculus of Inductive
Constructions. The main difference from proof assistants based on
higher-order logic is the presence of dependent types. Coq also
features a rich tactic language Ltac, which allows to write
specialised proof automation tactics. Some standard automation tactics
already available in Coq are:
\begin{itemize}
\item \texttt{intuition}: implements a decision procedure for
  intuitionistic propositional calculus based on the contraction-free
  sequent calculi LJT* of Roy Dyckhoff.
\item \texttt{firstorder}: extends \texttt{intuition} to a proof
  search tactic for first-order intuitionistic logic.
\item \texttt{auto} and \texttt{eauto}: implement a Prolog-like
  backward proof search procedure.
\end{itemize}

The CoqHammer~\cite{lcck-jar18,CzajkaKaliszyk2016} plugin extends Coq
automation by a number of other useful and generally more powerful
tactics similarly to that available in Isabelle~\cite{jbdgckdkju-jar-mash16}. Its main tactic \texttt{hammer} combines machine learning and
automated reasoning techniques to discharge goals automatically. It
works in three phases:
\begin{enumerate}
\item {\bf Premise selection} uses machine learning techniques to
  choose a subset of the accessible lemmas that are likely useful for
  the goal.
\item {\bf Translation} of the goal and the preselected
  lemmas to the input formats of first-order automated theorem
  provers~(ATPs) such as Vampire~\cite{KovacsVoronkov2013} or
  Eprover~\cite{Schulz2013}, and running the~ATPs on the translations.
\item {\bf Reconstruction} uses the information obtained from a
  successful~ATP run to re-prove the goal in the logic of Coq. Upon
  success the \texttt{hammer} tactic should be replaced with the
  reconstruction tactic displayed in the response window. The success
  of the reconstruction tactic does not depend on any time limits nor
  external ATPs, therefore it is machine-independent.
\end{enumerate}
 The CoqHammer tool provides various reconstruction tactics.
% which upon success are output by
%  the \texttt{hammer} tactic in the response window.
Among others, the tactics \texttt{hobvious} and \texttt{hsimple}
perform proof search via the \texttt{yelles} tactic (see the last item below)
using the information returned from the successful ATP runs
after a constant unfolding and hypothesis simplification.
Also,
%to the \texttt{hammer} tactic which aims at proving relatively
%simple goals using available lemmas in the context, 
CoqHammer comes with
tactics written entirely in Ltac. These tactics do not depend
on any external
tool, and are not informed about available lemmas in the context:
\begin{itemize}
\item \texttt{sauto} -- a ``super'' version of the standard Coq tactics
  \texttt{auto} and \texttt{intuition}. It tries to simplify the goal
  and possibly solve it without performing much of actual proof search
  beyond what \texttt{intuition} already does. It is designed in such
  a way as to terminate in a short time in most circumstances.  One can
  customize it by adding rewrite hints to the
  \texttt{yhints} database.
\item \texttt{scrush} -- essentially a combination of \texttt{sauto}
  and \texttt{ycrush}. The \texttt{ycrush} tactic tries various
  heuristics and performs some limited proof search.  Usually stronger
  than \texttt{sauto}, but may take a long time if it cannot find a
  proof. In contrast to \texttt{sauto}, \texttt{ycrush} does not
  perform rewriting using the hints in the \texttt{yhints}
  database. One commonly uses \texttt{ycrush} after \texttt{sauto}.
% for
%  goals which \texttt{sauto} could not solve.
\item \texttt{yelles n} -- 
performs proof search up to depth n; slow for n larger than 3-4.
\end{itemize}

\section{Case Studies}
\label{castd}
In this section, we illustrate a set of goals that are discharged using the Coq automation
techniques, presented in Section~\ref{coqa}, together
with a comparison to their original versions, in an Isabelle/HOL formalization,  as presented
in the Concrete Semantics book. Notice that the examples in this section are given broadly, with no
background details. The point to emphasize here is that we can
actually achieve a similar brevity of the formal text in terms of proof lengths using proof
automation in Coq. The examples are given in code snippets that have Coq text on the
left and Isabelle/HOL text on the right side of the minipages.

Note also that we translated thelemma statements into Coq directly from
Isabelle/HOL theory files, and proved them using mostly the standard tactics
coming with CoqHammer, with only minimal use of more sophisticated
custom Ltac tactics, and practically no hints from Coq hint
databases. Therefore the translation is not quite automatic but
fairly straightforward.

The example given in the below code snippet comes from the Hoare Logic. Leaving the technical details aside,
it basically says that a precondition \texttt{\{P\}} of some Hoare triple can be strengthened
into  \texttt{\{P'\}} if  \texttt{\{P'\}} entails  \texttt{\{P\}}. This is actually one of the corollaries
of  the consequence (called \texttt{conseq} in our Coq formalization) rule of Hoare Logic.
Notice that, in this snippet, \texttt{hoaret} is the Coq inductive predicate representing Hoare triples
which corresponds to the notation ``$\mathtt{\vdash}_t$'' on Isabelle/HOL side.
%{\vskip -0.4cm}
%\begin{figure}[h]\begin{figure}[h]
\begin{minipage}[h]{0.6\textwidth}
\begin{tcb}[\scriptsize]{Coq}
Lemma strengthen_pre:
forall (P P' Q: assn) c, (entails P' P)
-> hoaret P c Q -> hoaret P' c Q.
Proof. hobvious Empty (@conseq) (@entails) Qed.
\end{tcb}
\end{minipage}
\begin{minipage}[h]{0.35\linewidth}
\setlength{\fboxsep}{1pt}
\scriptsize
\Snippet{strengthenpre}
\end{minipage}
%{\vskip -0.15cm}
%\caption{H. L. for total correctness: strengthening preconditions.}
%\label{hlsp}
%\end{figure}

Upon a call, the CoqHammer tool gets a proof returned by one of the employed ATPs, and
discharges the goal using its reconstruction tactic \texttt{hobvious} parametrized with the empty set
of hypotheses from the goal context, the rule \texttt{conseq} and the definition \texttt{entails}.
Indeed, this is very similar to what happens in Isabelle/HOL proof of the same fact. The proof is
simply made of a call to the \texttt{metis} tactic with the \texttt{conseq} rule as the argument. 

Another but slightly more complicated example that stems from the Hoare Logic (using the
same notation as the previous one) is given in the below code snippet. 
This lemma is a version of the partial correctness of the \texttt{while} rule
enriched with a measure function \texttt{f} which is supposed to decrease in each loop iteration
so as to guarantee the loop termination.
%\begin{figure*}[t]
%\centering
{\vskip -0.5cm}
\begin{minipage}[h]{0.47\textwidth}
\begin{tcb}[\scriptsize]{Coq}
Lemma While_fun: forall b P Q c
(f: state -> nat), (forall n: nat, hoaret 
(fun s => P s /\ bval s b = true/\ 
n = f s) c (fun  s => P s /\ f s < n))
-> hoaret P (While b c)  
(fun s => P s /\  bval s b = false). 
Proof. pose While; pose conseq; 
unfold entails in *; yelles 3. Qed.
\end{tcb}
\end{minipage}
\begin{minipage}[h]{0.48\textwidth}
\setlength{\fboxsep}{1pt}\scriptsize
\Snippet{whilefun}
\end{minipage}
{\vskip -0.25cm}
%\caption{H. L. for total correctness: partial \texttt{While} rule with a measure function.}
%\label{hlw}
%\end{figure*}
The Coq proof is found by the Ltac implemented tactic \texttt{yelles} which performs
a proof search until a user specified depth has been reached. In our concrete example, we give it
some guidance by using the primitive Coq tactic \texttt{pose} with \texttt{while} and \texttt{conseq} rules
as arguments, adding them to the context (or simply generalizing them), together with unfolding the definition
of \texttt{entails}. This way, the tactic finds a proof at the proof search depth 3. Isabelle/HOL proof of the same
statement follows similar lines. It uses a simplification of the \texttt{while} rule with the measure
function being  $\mathtt{\lambda s\,n.\,n\,=\,f\,s}$. Just notice that our Coq tactic \texttt{yelles} is
clever enough on this goal to find the measure function automatically.

A third example is about semantics of the \texttt{IMP} language. 
The lemma shown in the below snippet states that one can deduce the big-step semantics
of any terminating \texttt{IMP} program from its small-step semantics.
Observe that, in this snippet, the Coq notations ``$\mathtt{\Longrightarrow}$'' and
``$\mathtt{\longrightarrow\texttt{*}}$'' respectively represent the inductive predicates
for the (transitive closure of) \texttt{IMP} big-step and small-step semantics.
The single difference on Isabelle/HOL side is that we have ``$\mathtt{\Rightarrow}$''
standing for big-step semantics.
{\vskip -0.35cm}
%\begin{figure}[t]
%\centering
\begin{minipage}[h]{0.53\linewidth}
\begin{tcb}[\scriptsize]{Coq}
Lemma lem_small_to_big: forall p s,
p -->* (Skip, s) -> p ==> s.
Proof. enough (forall p p', p -->* p' -> 
forall s, p' = (Skip, s) -> p ==> s) by scrush.
intros p p' H. induction H; sauto.
hsimple AllHyps (@lem_small1_big_continue)
Empty. Qed.
\end{tcb}
\end{minipage}
\begin{minipage}[h]{0.42\linewidth}
\setlength{\fboxsep}{1pt}\scriptsize
\Snippet{smalltobig}
\end{minipage}
{\vskip -0.2cm}
%\caption{standard \texttt{IMP}: small step implies big step semantics.}
%\label{impsib}
%\end{figure}

The Coq proof of this lemma proceeds by an induction on the (transitive closure of) small-step semantics
after introducing an helper statement (asserted by the 
pure Coq tactic \texttt{enough} and proven by the Ltac tactic \texttt{scrush}) into the goal context. 
Then, it calls the Ltac tactic \texttt{sauto} to
do some preprocessing for the CoqHammer call.
The base case $\mathtt{p\,=\,(Skip,\,s)}$  is trivially
solved by \texttt{sauto}. For the inductive case, namely 
$\mathtt{\forall s,\, p'\,=\,(Skip,\,s)\rightarrow p'\,\texttt{==>\,s}}$, we
call CoqHammer and get the goal solved by an application of the reconstruction tactic 
\texttt{hsimple} which uses all hypotheses
in the goal context (that's why we introduce a new one at the
beginning) 
and the helper lemma called
$\mathtt{lem\_small1\_big\_continue}$ with no definitions unfolded. 
This is again very similar to
the Isabelle/HOL proof of the fact in hand. The proof uses the induction principle on the transitive closure
of the small-step \texttt{IMP} semantics and then applies the helper lemma
$\mathtt{lem\_small1\_big\_continue}$.
%
%We provide three more examples in the Appendix~\ref{appendix}
%that we think interesting and possibly help the process of reviewing.
%
%We have already shown three proofs examples in the paper,
%all are available in the attached formalization but
Below we give three more examples that
we think interesting in the sense that all cases appear on Coq side are discharged
fully automatically. And the text size is fairly close to the one of Isabelle/HOL. 
%This lemma states that if \texttt{n} step execution
%of a compiler using a list of instructions \texttt{P} on a configuration
%(basically made of a stack and a state) $\mathtt{c_1}$ into $\mathtt{c_2}$, then there exists
%a transitive closure of execution step using the same 
%\begin{figure}[t]
%\centering
{\vskip -0.35cm}
\begin{minipage}[h]{0.48\linewidth}
\begin{tcb}[\scriptsize]{Coq}
Lemma exec_n_exec: forall n P c1 c2,
exec_n P c1 n c2 -> exec P c1 c2.
Proof. induction n; intros; destruct H.
- scrush. 
- pose @star_step;
hobvious (@H, @IHn)(@Star.star_step)
(@Compiler.exec). Qed.

Lemma exec_exec_n: forall P c1 c2,
exec P c1 c2 ->exists n, exec_n P c1 n c2.
Proof. intros; induction H.
- exists O; scrush. 
- pose exec_Suc; scrush. Qed.

Lemma exec_eq_exec_n: forall P c1 c2, 
exec P c1 c2 *-> exists n, exec_n P c1 n c2.
Proof. pose exec_exec_n; 
pose exec_n_exec; scrush. Qed.
\end{tcb}
\end{minipage}
%{\hskip -1.25cm}
\begin{minipage}[h]{0.45\linewidth}
\setlength{\fboxsep}{1pt}\scriptsize
\Snippet{execexec}
\end{minipage}
{\vskip -0.15cm}

The main lemma $\mathtt{exec\_eq\_exec\_n}$, using the other two above,
 is broadly about the symbolic compilation of
 \texttt{IMP} programs into a low level language based on a stack machine. 
It specifically says that one can speak about the \texttt{n} step instruction executions
instead of reflexive transitive closure of single step executions.
Note that the Coq predicates $\mathtt{exec\_n}$ and \texttt{exec} are respectively standing for
\texttt{n} step instruction, and transitive closure of single step instruction.
These are denoted as ``$\mathtt{\_\vdash\_\rightarrow*\,\_}$'' and 
``$\mathtt{\_\vdash\_\rightarrow^{\wedge}n\,\_}$'' in Isabelle/HOL text respectively.
%in Figure~\ref{imptc}.
%
The Coq proof from left to right
($\mathtt{exec\_n\_exec}$) of the equivalence is based on an induction
over \texttt{n} and the other direction ($\mathtt{exec\_exec\_n}$) is based on an
induction over transitive closure of single step
executions. The base cases of both induction steps are trivially
solved by the tactic \texttt{scrush}. The inductive
case of the former proof is a CoqHammer call which discharges
the goal using the reconstruction tactic \texttt{hobvious}.  It
uses two hypotheses (\texttt{H} and the induction hypothesis
\texttt{IHn}) from the goal context with the lemma called $\texttt{star\_step}$
(coming from \texttt{Star.v}), and the
definition \texttt{exec}.
The inductive case of the latter is just made of an \texttt{scrush} application with a guide reminding
that the goal is a variant of the lemma $\mathtt{exec\_Suc}$. 

Again, these proofs follow very similar lines with those of Isabelle/HOL of the same facts. The proof of
$\mathtt{exec\_n\_exec}$ induces on \texttt{n} and uses the definition $\mathtt{star\_step}$ to
discarge the goal. Similarly, $\mathtt{exec\_exec\_n}$ is proven by an induction on the transitive
closure of single step executions followed by the application of the $\mathtt{exec\_Suc}$ fact.

%\Snippet{strengthenpre}
%Multiple examples side-by-side.
%
%Which ones? What to write about them?
%
%Generation from Isabelle example:

\section{Conclusion}
\label{concl}
We have reproven 101 lemmas from the Isabelle/HOL
theories \texttt{Star}, \texttt{AExp}, \texttt{BExp}, \texttt{ASM}, \texttt{Com}, 
\texttt{Big$\_$Step},  \texttt{Hoare}, \texttt{Small$\_$Step},
\texttt{Compiler} and \texttt{Compiler2} in Coq; heavily using the automation techniques
described in previous sections. 
%{\vskip -1cm}
\begin{center}
\begin{tabular}{ |p{2.1cm}||p{1.6cm}|p{1.75cm}|p{1.75cm}|p{2.75cm}| p{1.5cm}|}
% \hline
 \hline
& \# of lines &\# of words&\# of tactics&\# of hammer calls&time(secs)\\
 \hline
 Isabelle/HOL   & 2806    & 11278 & 544 & not verifiable&31\\
 Coq &   3493  & 19292   & 1190 & 468&149\\
 \hline
\end{tabular}
\end{center}
%{\vskip -0.2cm}
\noindent
%The above table shows the number of lines, words, tactic applications
%(coumpoud tactic calls counted one) and succesful hammer calls
%used in proof texts.
As shown in the above table, the number of Coq tactics we used to get the same
lemmas proven is almost twice in number, as opposed to Isabelle/HOL,
but about half of which benefits from the
automation techniques that CoqHammer comes with. This can be seen
as an improvement given that the Isabelle/HOL tactics are more compound than
the ``simple'' Coq tactics.

The \texttt{coqc 8.7.2} needs 149 seconds to compile the translated source
and Isabelle 2017 needs 31
seconds to build the corresponding theories on an Intel Core i7-7600U machine. We attribute
the difference mostly to the fact that all used Isabelle tactics are written in ML, while
most Coq ones use Ltac.

We plan to build on this work by proving more lemmas coming from different theories
of the book and by improving the level of automation, thus decreasing
 the number of words, in the already proven goals. 
%
% In total, we have performed XX calls to CoqHammer together with
%XX calls to the  automation tactics such as \texttt{scrush}, \texttt{yelles}, \texttt{sauto}
%and \texttt{omega}. 
%And in XX number of goals, we managed to compress our proof text close enough to
%the original text by XX percent. In the rest, even though proof texts are not as short as
%the original ones, we still managed to inject a heavy proof automation. 
%We plan to build on this work by proving more lemmas coming from different theories
%of the book and by improving the level of automation, thus shorthening the proof texts,
%in the already proven goals. 
Please see:

  \centerline{\url{https://github.com/lukaszcz/COQ-IMP}}
\noindent
for the proofs done so far. 
%Note that the sources compile with Coq version 8.7.2.
% Likely no need...

\paragraph{Acknowledgments}
This work has been supported by the Austrian Science Fund (FWF) grant
P26201, the European Research Council (ERC) grant no.~714034
\emph{SMART} and the Marie Sk{\l}odowska-Curie action \emph{InfTy},
program H2020-MSCA-IF-2015, number 704111. The first author was
supported entirely by the European Union's Horizon 2020 research and
innovation programme under the Marie Sk{\l}odowska-Curie grant
agreement number~704111.

\bibliographystyle{abbrv}
\bibliography{s}

\newpage

%\appendix 
%\section{Appendix}
%\label{appendix}

\end{document}